\documentstyle[12pt]{article}

\def\beq{\begin{equation}}
\def\eeq{\end{equation}}
\def\bea{\begin{eqnarray}}
\def\eea{\end{eqnarray}}
\def\bq{\begin{quote}}
\def\eq{\end{quote}}

\def\CQG{{\it Class.Quantum Grav.} }

\def\NP{{\it Nucl.Phys.} }

\def\PR{{\it Phys.Rev.} }

\begin{document}

\title{
DECOHERENCE OF HOMOGENEOUS AND ISOTROPIC METRICS\\
IN THE PRESENCE OF \\ MASSIVE VECTOR FIELDS\thanks{Talk presented at
the MG7 Conference, Stanford University,
July 24-29, 1994.}}
\author{{\large\bf O. Bertolami}\thanks{On leave from Departamento de
F\'{\i}sica,
Instituto Superior T\'ecnico,
Av. Rovisco Pais, 1096 Lisboa, Portugal; e-mail: orfeu@vxcern.cern.ch}\\
Theory Division, CERN, \\ CH - 1211 Geneva 23,  Switzerland
\and
{\large\bf P.V. Moniz}\thanks{Work supported by a Human Capital and Mobility
Fellowship from the European
Union (Contract ERBCHBICT930781); e-mail: prlvm10@amtp.cam.ac.uk}\\
University of Cambridge, DAMTP, \\   Cambridge, CB3 9EW, UK}

\date{.}

\maketitle

\begin{abstract}
Retrieval of classical behaviour in quantum cosmology is usually discussed in
the framework of minisuperspace models in the presence of scalar fields
together with the inhomogeneous modes either of the gravitational or of the
scalar fields. In this work we propose alternatively a  model
where the scalar field is
replaced by a massive vector field with global $U(1)$ or SO(3) symmetries.
\end{abstract}

\medskip

The emergence of the classical properties from the quantum mechanics formalism
is still largely an open problem. Some progress has, however, been achieved
through the so-called decoherence approach.  On
fairly general grounds, decoherence can be regarded as a procedure
where one considers the
system under study to be part of a more complex world and which interacts with
other subsystems, usually referred to as ``environment". This interaction leads
to the suppression of the quantum interference effects.

These ideas have been developed with some depth in the
context of minisuperspace models in quantum cosmology (see \cite{OBPM} and
references therein).
Most of the literature concerning the emergence of classical behaviour from
quantum cosmological minisuperspace models considers scalar fields
and as environment the inhomogeneous modes either of the
gravitational or of the scalar fields.
We propose alternatively a model where the self-interacting scalar field is
replaced by a massive vector field with U(1) or SO(3) global symmetries
\cite{OBPM}.
Preliminary work on the  system
with $SO(3)$ non-Abelian global symmetry, whose classical cosmology has been
studied in Ref. \cite{PP},
shows that
the ingredients necessary for the process of decoherence to take place are
present \cite{SS}. Notice that the presence of a mass term
is an essential feature as this breaks the
conformal symmetry of the spin-1 field action which leads to a Wheeler-DeWitt
equation where gravitational and matter degrees of freedom decouple
\cite{LLL}.

The action of our model consists of a Proca field coupled with gravity
\cite{PP}:
\beq
S = \int d^4x \sqrt{-g}\left[ {1\over 2k^2} (R - 2\Lambda) + {1\over 4e^2}
{\rm Tr}({\boldmath  F}_{\mu\nu}{\boldmath F}^{\mu\nu}) + {1\over 2}~m^2
{\rm Tr}({\boldmath A}_{\mu} {\boldmath A}^{\mu})\right]
-{1\over k^2} \int_{\partial M} d^3x \sqrt{h} K ,
\label{1}
\eeq
where $k^2 = 8\pi M^{-2}_P$, $M_P$ being the Planck mass, $e$ is a gauge
coupling constant, $m$ the mass of the Proca field and
with $h_{ij} (i,j = 1,2,3)$ being the induced metric on the three-dimensional
boundary $\partial M$ of $M$, $h = \det (h_{ij})$ and $K = K^{\mu}_\mu$ is the
trace of the second fundamental form on $\partial M$.

In quantum cosmology one is concerned with spatially compact topologies
 and we will consider
the Friedmann-Robertson-Walker (FRW)
ansatz for  the  ${\bf R}\times S^3$  geometry
\beq
ds^2 = \sigma^2 a^2(\eta) \left[ -N(\eta)^2 d\eta^2 + \sum^3_{i=1}
\omega^i\omega^i\right]~,
\label{2}
\eeq
 where
 $\sigma^2 = 2/3\pi$, $\eta$  is the conformal time,
$N(\eta)$ and $a(\eta)$ being the lapse function
and the scale factor, respectively and $\omega^i$ are left-invariant
one-forms in $SU(2)\simeq$ $S^3$.

Aiming to obtain solutions of the Wheeler-DeWitt equation satisfied by the wave
function $\Psi [h_{ij},A^{(a)}_{\mu}]$
we expand the spatial metric as:
\beq
h_{ij} = \sigma^2a^2 ~~(\Omega_{ij} +\epsilon_{ij})
\label{7}
\eeq
with $\Omega_{ij}$ being the metric on the unit $S^3$ and $\epsilon_{ij}$ a
perturbation which can be expanded in in scalar harmonics
${\cal D}^{J{\phantom J}M}_{{\phantom J}N}(g)$,
which are the usual $(2J+1)$-dimensional $SU(2)$
matrix representation, and spin-2 hyperspherical harmonics $Y^{2~LJ}_{m~MN}(g)$
on $S^3$ \cite{MM}.

The massive vector field
\begin{equation}
{\boldmath A} = A_m^{ab}\omega_{s}^m {\cal T}_{ab} =
 A_m^{ab} \sigma^m_i \omega^i {\cal T}_{ab}~,
\label{vec}
\end{equation}
where $\omega_{s}^m$ denote the one-forms in a spherical basis with
$m=0,\pm 1$, $\sigma^m_i$ denotes a 3 x 3 matrix and
${\cal T}_{ab}$ are the $SO(3)$ group generators,
can be expanded in spin-1 hyperspherical harmonics as \cite{OBPM},\cite{TT}:
\bea
A_0(\eta,x^j) &=& \sum_{JMN} \alpha^{abJM}_{\phantom{^{JM}} N}(\eta) {\cal
D}^{J{\phantom J}M}_{{\phantom J}N} (g) {\cal T}_{ab}
= 0 + \sum_{J'M'N'} \alpha^{abJ'M'}_{\phantom{^{J'M'}} N'}(\eta) {\cal
D}^{J'{\phantom J'}M'}_{{\phantom J'}N'} (g) {\cal T}_{ab},
\label{11a} \\
A_i(\eta,x^j) &=& \sum_{LJNM} \beta^{abMN}_{LJ} (\eta)
{}~Y^{1LJ}_{mNM}(g) \sigma^m_i~{\cal T}_{ab} \nonumber \\
 & = & {1 \over 2}\left[1 + \sqrt{{{2\overline{\alpha}}
\over
{3\pi}}} \chi(\eta)\right]\epsilon_{aib}{\cal T}_{ab}
+ \sum_{L'J'N'M'} \beta^{abM'N'}_{L'J'} (\eta)
{}~Y^{1L'J'}_{mN'M'}(g) \sigma^m_i~{\cal T}_{ab},
\label{11b}
\eea
where $\overline{\alpha} = e^2/4\pi$ and $\chi(\eta)$ a
 time-dependent scalar function.
$A_0$ is a scalar on each fixed time
hypersurface, such that it can be expanded in scalar harmonics
${\cal D}^{J{\phantom J}M}_{{\phantom J}N}(g)$.
The expansion of $A_i$ is performed in terms of the spin-1 spinor
hyperspherical harmonics, $Y^{1~LJ}_{m~MN}(g)$. Longitudinal and the
transversal harmonics correspond to $L = J$ and $L-J = \pm 1$, respectively.
The rhs of eqs.
(\ref{11a}) and (\ref{11b})
correspond to a decomposition in homogeneous and
inhomogeneous modes. For this decomposition one has used
the ansatz for the homogeneous modes
of the vector field which is compatible with the FRW geometry as discussed in
Ref. \cite{PP}.

 From action (\ref{1})
one can work out the effective Hamiltonian density obtained from the
substitution of the
expansions (\ref{2})-(\ref{11b}). To second order in the coefficients of the
expansions and in all orders in $a$, one obtains
the following effective Hamiltonian density for the system with SO(3) global
symmetry (for the Abelian case one drops the last four terms) \cite{OBPM}:
\bea
{\cal H}^{{\rm eff}} &=& -\frac{1}{ 2M_P^2} \pi^2_a + M_P^2\left( - a^2 +
{4\Lambda\over 9\pi }~a^4
\right) \nonumber
+ \sum_{J,L}~{4\over 3\pi}~m^2~a^2~\beta^{abNM}_{LJ}
{}~\beta^{abLJ}_{NM} \nonumber \\
&& + \sum_{\vert J-L\vert =1} \left[ \overline{\alpha}\pi
\Pi_{\beta^{abLJ}_{NM}}
\Pi_{\beta^{abLJ}_{NM}} + \beta^{abNM}_{LJ} \beta^{abLJ}_{NM}
 (L+J+1)^2\right]
\nonumber \\
&& + \sum_J \left\{
\overline{\alpha}\pi + \left[(-1)^{4J} \left({16\pi^2 J(J+1)\over 2J+1}\right)
{}~{3\pi
\over 4m^2}\right]~~
\left[ 1 + {1\over a(t)}\right] \right\}\Pi_{\beta^{abJJ}_{NM}}
\Pi_{\beta^{abJJ}_{NM}} \nonumber \\
&& + ~\pi^2_{\chi} + {\overline{\alpha}\over 3\pi} \left[ \chi^2  - {3\pi\over
2\overline{\alpha}}\right]^2  \nonumber
+ \sum_{J,L} {4\over \overline{\alpha}\pi}~\left[ 1 +
\sqrt{{2\overline{\alpha}\over 3\pi}} \chi
\right]^2~\beta^{abNM}_{LJ} ~\beta^{abLJ}_{NM} \nonumber \\
&& + 4\pi a^2 m^2\left[ 1 + \sqrt{{2\overline{\alpha}\over
3\pi}}\chi\right]^2,
\label{12}
\eea
where the canonical conjugate momenta of the dynamical variables are given by
\beq
\pi_{a} = {\partial {\cal L}^{{\rm eff}}\over\partial\dot a}
= -\frac{\dot a}{N}~,~~
\pi_{\chi} =
{\partial {\cal L}^{{\rm eff}}\over \partial \dot\chi} = \frac{\dot\chi}{N}~,~~
\pi_{\beta^{abNM}_{LJ}} = {\partial {\cal L}^{{\rm eff}}
\over\partial\dot\beta^{abNM}_{LJ}}
= \frac{\dot\beta^{abNM}_{LJ}}{2N\pi\overline{\alpha}}~,
\label{13a}
\eeq

\beq
\pi_{\beta^{abNM}_{JJ}} = {\partial {\cal L}^{{\rm eff}}
\over\partial\dot\beta^{abNM}_{JJ}}
= \frac{\dot\beta^{abNM}_{JJ}}{2N\pi\overline{\alpha}} -
\frac{1}{2\pi\overline{\alpha}}\frac{a}{N} \alpha^{abJM}_{\phantom{^{JM}} N}
(-1)^{2J} \sqrt{{16\pi^2 J(J+1)\over 2J+1}}~,
\label{13b}
\eeq
${\cal L}^{{\rm eff}}$ denoting the effective Lagrangian density
arising from (\ref{1}) and the dots representing derivatives with respect to
the conformal time.

The Hamiltonian constraint, ${\cal H}^{{\rm eff}} = 0$, gives origin to the
Wheeler-DeWitt equation after promoting the canonical conjugate momenta
(\ref{13a}),(\ref{13b}) into operators:
\beq
\pi_{a} = -i{\partial\over\partial a}~,~~\pi_{\chi} =
-i{\partial\over\partial
\chi}~,~~ \pi_{\beta^{abNM}_{LJ(JJ)}} = -i{\partial\over\partial
\beta^{abLJ(JJ)}_{NM}}~,
{}~~\pi_{a}^2 = - a^{-P}{\partial\over\partial a}~
\left( a^P{\partial\over\partial a}\right)~,
\label{14}
\eeq
where in (\ref{14}) the last substitution parametrizes the operator order
ambiguity with $p$ being a real constant.
The Wheeler-DeWitt equation is obtained imposing that the Hamiltonian operator
annihilates the wave function
$\Psi [a,\beta^{abNM}_{LJ},\beta^{abNM}_{JJ},\chi]$.

A solution of the Wheeler-DeWitt equation
which corresponds to a classical behaviour
of its variables on some region of minisuperspace will have an
oscillatory WKB form as
\begin{equation}
\Psi[a,A^{ab}_\mu] = e^{iM_P^2 S(a)} C(a) \psi (a,A^{ab}_\mu)~.
\label{WKB}
\end{equation}

In order to make predictions concerning the behaviour of the scale factor, $a$,
for the U(1) case (see Ref. \cite{OBPM} for the discussion of the
non-abelian case with SO(3) global symmetry)
one uses a coarse-grained description of the system
working out the reduced density matrix associated to (\ref{WKB})
\begin{equation}
\rho_R = \sum_{n,n'} e^{iM_P^2[S_{(n)}(a_1) - S_{(n')}(a_2)]}
C_{(n)}(a_1)C_{(n')}(a_2) {\cal I}_{n,n'}(a_2,a_1),
\label{red}
\end{equation}
where
\begin{equation}
{\cal I}_{n,n'}(a_2,a_1) = \int \psi_{(n')} (a_2,A^{ab}_{\mu}) \psi^*_{(n)}
(a_1,A^{ab}_{\mu}) d[A^{ab}_{\mu}]~.
\label{infl}
\end{equation}
The subindex $(n)$ labels the WKB branches. The term
${\cal I}_{n,n'}(a_2,a_1)$ contains
the environment influence on the system.

The decoherence process is sucessful if the non-diagonal terms $(n\neq n')$ in
(\ref{red}) are vanishingly small. Hence, there will be no quantum interference
between alternative histories if $ {\cal I}_{n,n'} \propto \delta_{n,n'}$.
Once that is achieved one can analyse the correlations in each classical
branch ($n = n'$). This can be done by looking at the reduced density matrix
or the to corresponding Wigner functional:
\begin{equation}
F_{W,(n)} (a,\pi_a) = \int_{-\infty}^{+\infty} d\Delta
[S'_{(n)}(a_1) S'_{(n)}(a_2)]^{-\frac{1}{2}} e^{-2i\pi_a\Delta}
e^{iM_P^2[S_{(n)}(a_1) - S_{(n)}(a_2)]} {\cal I}_{n,n}(a_2,a_1)
\label{wigf}
\end{equation}
where $\Delta = \frac{a_1 - a_2}{2}$.
A correlation among variables will correspond to a strong peak
about a classical trajectory in the phase space. The decoherence process
is necessary as the Wigner function associated to
(\ref{red}) does not have a single sharp peak even for a WKB Wigner function as
(\ref{wigf}); such a peak (and a clearly classical WKB evolution) is found
only among the $n=n'$ terms.
If the conditions
to achieve an effective diagonalization of (\ref{wigf}) are met then the
interference between the different classical behaviours is {\em also}
highly suppressed.
Furthermore, ${\cal I}_{n,n'}(a_2,a_1)$ will be damped for
$|a_2 - a_1|\gg 1$ and the
reduced density matrix associated with (\ref{wigf}) will be diagonal
with respect to the variables $a$ \cite{OBPM}.

\vspace{1cm}

{\bf Acknowledgments}

\vspace{0.3cm}

The authors are very grateful to U. Ancarani,  H.F. Dowker, J.S. Dowker,
L. Garay, B.L. Hu, C. Kiefer, J. Mour\~ao and J.P. Paz for valuable
discussions.

\end{document}